\documentclass[twocolumn]{jpsj2}

\usepackage{amsmath}
\usepackage{amsfonts}
\usepackage{amssymb}
\usepackage{txfonts}
\usepackage{bm}
\usepackage{tabularx}
\usepackage{graphicx,color}
\def\Vec#1{\bm{#1}}

\def\runauthor{Online-Journal Subcommittee}

\title{%
Spatial development of superconductivity in the Sr$_2$RuO$_4$-Ru eutectic system
}
\author{%
Shunichiro Kittaka$^{1,\thanks{E-mail: kittaka@scphys.kyoto-u.ac.jp}}$,
Taketomo Nakamura$^1$, Hiroshi Yaguchi$^{1,2}$, Shingo Yonezawa$^1$, and Yoshiteru Maeno$^1$
}
\inst{%
$^1$Department of Physics, Graduate School of Science, Kyoto University, Kyoto 606-8502, Japan\\
$^2$Department of Physics, Faculty of Science and Technology, Tokyo University of Science, Noda 278-8510, Japan
}
\recdate{\today}
\abst{%
We have clarified how the enhanced superconductivity, often referred to as the 3-K phase superconductivity, 
develops in the Sr$_2$RuO$_4$-Ru eutectic system.
From the detailed ac and dc susceptibility measurements on well-characterized crystals, we revealed 
strongly anisotropic shielding, governed by the direction of the screening current dominated within the RuO$_2$ plane rather than by
the orientation of the Ru lamellae.
The onset temperature of the 3-K phase superconductivity probed by diamagnetic screening is as high as 3.5~K. 
The temperature dependence of the diamagnetic shielding above around 2~K 
is well ascribed by the interfacial screening around each Ru lamella.
Below around 2~K, the rapid development of the shielding fraction as well as its peculiar response to ac and dc magnetic fields 
are explained by the formation of the Josephson network consisting of inter-lamellar supercurrents.
}
\kword{%
Sr$_2$RuO$_4$, ruthenate, eutectic crystal, proximity effect, diamagnetic shielding fraction
}

\begin{document}

\maketitle
\section{Introduction}
The Sr$_2$RuO$_4$-Ru eutectic system, in which Ru lamellae are regularly embedded with a stripe pattern in a Sr$_2$RuO$_4$ single crystal, 
is fascinating because the superconducting transition temperature $T_\mathrm{c}$ is largely enhanced over those of Sr$_2$RuO$_4$ and Ru.
Pure Sr$_2$RuO$_4$ exhibits superconductivity at 1.5~K and is believed to be a spin-triplet $p$-wave superconductor with the vector order parameter $\Vec{d}(k)=\Hat{z}\Delta_0(k_x+ik_y)$,\cite{Maeno1994Nature,Mackenzie2003RMP}
and pure Ru metal is an $s$-wave superconductor with $T_\mathrm{c}=0.49$~K. 
Surprisingly, for Sr$_2$RuO$_4$-Ru eutectic crystals, 
a broad superconducting transition with an onset temperature of nearly 3~K has been observed through resistivity $\rho$ and ac susceptibility $\chi_\mathrm{ac}$ measurements.\cite{Maeno1998PRL}
The Sr$_2$RuO$_4$-Ru eutectic system with the enhanced $T_\mathrm{c}$ is referred to as the ``3-K phase".

Many interesting properties of the 3-K phase have been revealed, although the mechanism of the enhancement of $T_\mathrm{c}$ is still unknown.
First, it has been proposed that the 3-K phase superconductivity occurs at the interface between Sr$_2$RuO$_4$ and Ru.\cite{Maeno1998PRL}
Indeed, only a very tiny hump was observed in the specific heat.\cite{Yaguchi2003PRB}
Secondly, the 3-K phase superconductivity is closely related to the triplet pairing of Sr$_2$RuO$_4$.
For the 3-K phase, the upper critical field $H_\mathrm{c2}$ determined from resistance measurements has larger values for $H \parallel ab$ than for $H \parallel c$
($H_{\mathrm{c2} \parallel ab} / H_{\mathrm{c2} \parallel c} \sim 3.5$ at 0~K),\cite{Ando1999JPSJ}
which is the same tendency as $H_\mathrm{c2}$ for pure Sr$_2$RuO$_4$ ($H_{\mathrm{c2} \parallel ab} / H_{\mathrm{c2} \parallel c} \sim 20$ at 0~K) \cite{Deguchi2002JPSJ}.
In addition, tunneling measurements at the interface between Sr$_2$RuO$_4$ and Ru microinclusion
have revealed the presence of a zero bias conductance peak,\cite{Mao2001PRL,Kawamura2005JPSJ}
which is a hallmark of an unconventional superconductivity.\cite{Tanaka1995PRL} 
Very recently, Kambara $et$ $al$. observed an unusual hysteresis in the $I$-$V$ characteristics of micro-fabricated channels of the 3-K phase,\cite{Kambara2008PRL}
which indicates the existence of internal degrees of freedom in the superconducting order parameter.
These results suggest that the 3-K phase superconductivity is unconventional and sustained in the Sr$_2$RuO$_4$ part of the interface rather than in the Ru part.
Thirdly, the temperature dependence of $H_\mathrm{c2}$ in the 3-K phase is qualitatively different from that of Sr$_2$RuO$_4$.
Especially, an unusual upturn curvature in $H_\mathrm{c2}$ is observed at temperatures below 2~K for $H \parallel c$.\cite{Yaguchi2003PRB}

In order to explain these unusual features of the 3-K phase, 
Sigrist and Monien constructed a phenomenological theory 
which assumes a spin-triplet superconductivity occurring at the interface between Sr$_2$RuO$_4$ and Ru metal.\cite{Sigrist2001JPSJ}
They proposed that one of the two components of the superconducting order parameter, the component parallel to the interface, 
is stabilized at $T_\mathrm{c}$ of the 3-K phase
in zero field and that the other component with the relative phase of $\pi$/2 emerges at a slightly lower temperature.
Based on these experimental and theoretical works, 
the 3-K phase superconductivity is now believed to originate at the interface between Sr$_2$RuO$_4$ and Ru.

Although the typical distance between the nearest interfaces is approximately 10~$\muup$m,
zero-voltage current was observed at temperatures well above $T_\mathrm{c}$ of the bulk Sr$_2$RuO$_4$, $T_\mathrm{c-bulk}$.\cite{Hooper2004PRB} 
This observation suggests that the supercurrent flows between different interfaces.
In the present work, we performed detailed measurements of $\chi_\mathrm{ac}$ and dc magnetization $M$
as a more direct approach to examine the spatial development of the 3-K phase superconductivity. 
We revealed that the diamagnetic signal in $M$ for $H_\mathrm{dc} \parallel c$ becomes observable below approximately 3.5~K,
while it becomes observable only below 3~K for $H_\mathrm{dc} \perp c$.
The shielding fraction is also anisotropic with respect to the crystal structure of Sr$_2$RuO$_4$ 
rather than to the geometry of the interfaces.
The value of the shielding fraction is at most 1.2\% at 1.8~K, 
whereas the resistivity at this temperature becomes only 20\% of the normal-state value.\cite{Ando1999JPSJ,Kambara2008PRL}
Further decreasing temperature, we found that the diamagnetic shielding fraction rapidly increases and reaches nearly 100\% just above $T_\mathrm{c-bulk}$.
This large shielding fraction is more easily suppressed by the ac magnetic field $H_\mathrm{ac}$ than by the dc magnetic field $H_\mathrm{dc}$.
These ac and dc magnetic field responses as well as the large shielding fraction suggest that 
the Josephson network is formed among different interfaces 
with a long-range proximity effect in which the proximity length diverges toward $T_\mathrm{c-bulk}$.

\section{Experimental}

\begin{figure}
\begin{center}
\includegraphics[width=3.3in]{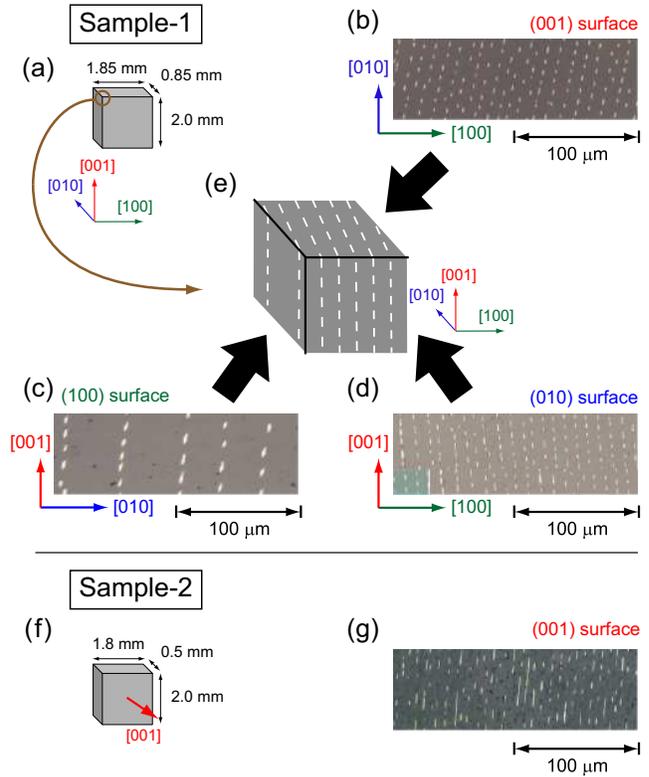}
\end{center}
\caption{(Color online) (a) Schematic illustration of Sample-1. 
Polarized-light optical microscopy images of the polished (b) (001), (c) (100), and (d) (010) surfaces of Sample-1 are shown.
The brighter and darker areas correspond to Ru and Sr$_2$RuO$_4$, respectively.
(e) Schematic illustration of the arrangement of Ru inclusions in Sample-1. 
(f) Schematic illustration of Sample-2. 
The in-plane axes of Sample-2 are not determined.
(g) Polarized-light optical microscopy image of the polished (001) surface of Sample-2.
}
\label{Sample1}
\end{figure}

The 3-K phase samples used in our study were grown using a floating zone method with an excess amount of Ru.\cite{Mao2000MRB}
We performed measurements on more than five 3-K phase samples from different batches and obtained qualitatively the same behavior.
In this paper, we focus on the results of two 3-K phase samples cut from different batches: Sample-1 and Sample-2.
As depicted in Fig.~\ref{Sample1} (a),
Sample-1, whose dimensions are $1.85 \times 2.0 \times 0.85$ mm$^3$, has a thin-plate shape parallel to one of the $ac$ planes of the tetragonal Sr$_2$RuO$_4$.
The crystal axes were determined from Laue pictures.
For convenience, we define one of the $a$ axes along the plate thickness as [010] and the other as [100].
Figures~\ref{Sample1} (b)-(d) show polarized-light optical microscope images of the polished (001), (100), and (010) surfaces of Sample-1.
The brighter and darker areas correspond to Ru and Sr$_2$RuO$_4$, respectively.
From these pictures, we found that 
the typical dimensions of the Ru inclusion are approximately 10 $\times$ 10 $\times$ 1 $\muup$m$^3$ with a thin-slab shape. 
As schematically illustrated in Fig.~\ref{Sample1} (e), these Ru lamellae are probably aligned nearly, but not exactly, parallel to the (100) plane.
The dimensions of Sample-2 are 1.8~mm $\times$ 2.0~mm in the $ab$ plane and 0.5~mm along the $c$ axis, as shown in Fig.~\ref{Sample1} (f).
A polarized-light optical microscope image of the polished (001) surface of Sample-2 is presented in Fig.~\ref{Sample1} (g).
Most regions of each surface of Sample-1 and Sample-2 show Ru stripe patterns similar to those shown in Fig.~\ref{Sample1}.
The total area of the Ru inclusions is estimated as high as 5\% of the area of the surface.

We measured $\chi_\mathrm{ac}=\chi^\prime-i \chi^{\prime \prime}$ by a mutual-inductance technique using a lock-in amplifier (LIA). 
The ac magnetic field $H_\mathrm{ac}$ was applied with a small hand-made coil (40~$\muup$T / mA). 
The dc magnetic field $H_\mathrm{dc}$ was generated by a 2-T magnet (Oxford Instruments), and was applied parallel to $H_\mathrm{ac}$. 
When we measured $\chi_\mathrm{ac}$ in zero dc field, 
we used a high-permeability-metal shield to exclude the geomagnetic field of about 50~$\muup$T.
The values of $\chi_\mathrm{ac}$ are obtained from the relation $\chi_\mathrm{ac}=iC_1V_\mathrm{LIA}/H_\mathrm{ac}+C_2$, 
where $C_1$ and $C_2$ are certain coefficients and $V_\mathrm{LIA}$($=V_x+iV_y$) is the read-out voltage of LIA.
We chose different values of $C_1$ and $C_2$ for each curve in Fig.~\ref{FD2}, whereas
we used the same values of $C_1$ and $C_2$ for all curves in Figs.~\ref{acac} and \ref{acdc}.
The samples were cooled down to 0.3~K with a $^3$He cryostat (Oxford Instruments, model Heliox VL). 
We measured $\chi_\mathrm{ac}$ at different frequencies ranging from 17~Hz to 10~kHz. 
Although $\chi_\mathrm{ac}$ of the 3-K phase superconductivity depends on frequency as reported in Ref.~14, 
we only present the data at 293~Hz since the behavior in $\chi_\mathrm{ac}$ with different frequencies are qualitatively the same. 
We measured $M$ with a SQUID magnetometer (Quantum Design, model MPMS) from 1.8~K to 4.2~K in the zero-field-cooling condition.

\section{Results}
\subsection{Anisotropy of the shielding fraction}

\begin{figure}
\begin{center}
\includegraphics[width=2.5in]{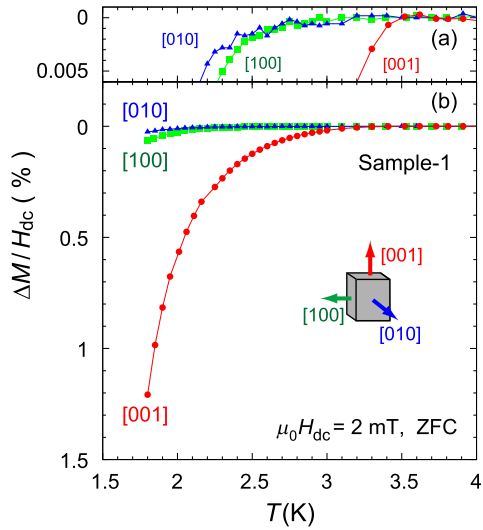}
\end{center}
\caption{(Color online) (a), (b) Temperature dependence of $\Delta M/H_\mathrm{dc}$ of Sample-1 in dc magnetic fields of 2~mT along the [100], [010], and [001] axes.
Figure (a) presents an enlarged view of $\Delta M/H_\mathrm{dc}$ near the onset.
}
\label{FD}
\end{figure}

We first present the result of the field-direction dependence of the 3-K phase superconductivity.
Figures~\ref{FD} (a) and \ref{FD} (b) show the temperature dependence of the dc shielding fraction $\Delta M/H_\mathrm{dc}$ in $\mu_0H_\mathrm{dc}$ of 2~mT along the [001], [100], and [010] axes.
In these figures, we normalize $M/H_\mathrm{dc}$ by the ideal value 
calculated for the full Meissner state without the demagnetization correction.
Remarkably, from Fig.~\ref{FD} (a), 
we found that $M$ for $H_\mathrm{dc} \parallel c$ starts to decrease with the onset temperature of 3.5~K, 
which is apparently higher than the onset temperature reported in the out-of-plane resistance $\rho_c$ measurements (at most 3~K) \cite{Ando1999JPSJ} 
but is comparable to that reported in the in-plane resistance measurement (above 3~K).\cite{Kambara2008PRL}
In contrast, the diamagnetic signal becomes observable only below 3~K for $H_\mathrm{dc} \parallel ab$, 
which is consistent with the onset temperature reported in the $\rho_c$ measurements. 
These results suggest that the 3-K phase superconductivity is highly two dimensional above 3~K.
As shown in Fig.~\ref{FD} (b), the anisotropy of the dc shielding fraction is also striking.
We found that the dc shielding fraction in the measurement temperature range strongly depends on the field direction with respect to
the Sr$_2$RuO$_4$ crystal axes rather than to the geometry of the Ru lamellae (cf. Figs.~\ref{Sample1} (a)-(e)).
The dc shielding fraction at 1.8~K is much larger for $H_\mathrm{dc} \parallel c$ than for $H_\mathrm{dc} \parallel ab$,
indicating that the screening current mainly flows within the $ab$ plane.
We note that the effect of the demagnetization factor of Sample-1 should be comparable between the fields along the [010] and [001] axes.

\begin{figure}
\begin{center}
\includegraphics[width=3in]{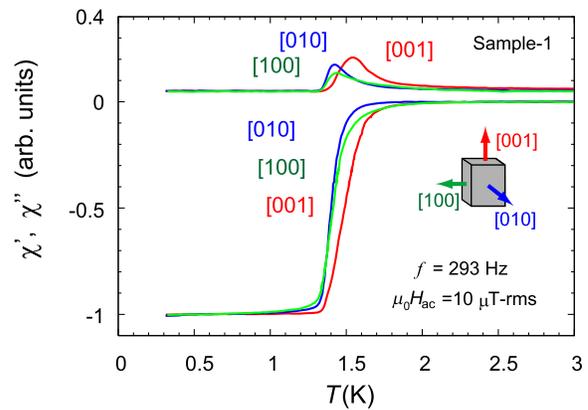}
\end{center}
\caption{(Color online) Temperature dependence of $\chi_\mathrm{ac}$ of Sample-1 in ac magnetic fields of 10 $\muup$T-rms along the [100], [010], and [001] axes in zero dc field.
The values of $C_1$ and $C_2$ for each curve are chosen so that $\chi^\prime(\mathrm{3~K})=0$ and $\chi^\prime(\mathrm{0.3~K})=-1$. 
}
\label{FD2}
\end{figure}

As shown in Fig.~\ref{FD2}, we also measured $\chi_\mathrm{ac}$ of Sample-1 with $\mu_0H_\mathrm{ac}$ of 10~$\muup$T-rms along the [100], [010], and [001] axes in zero dc field.
The values of $C_1$ and $C_2$ for each curve are chosen so that $\chi^\prime(\mathrm{3~K})=0$ and $\chi^\prime(\mathrm{0.3~K})=-1$. 
We note that these $\chi_\mathrm{ac}$ measurements cover lower temperatures than the $M$ measurements 
whereas its experimental resolution is relatively lower.
Although the onset between 2 and 3.5~K observed in $M$ is too small to be detected in $\chi_\mathrm{ac}$,
an anisotropic shielding fraction similar to that observed in the $M$ measurements was observed below 2~K. 
Near $T_\mathrm{c-bulk}$, the shielding fractions for the three field directions reach 100\% and the anisotropy apparently disappears.
This result indicates that the magnetic fluxes are excluded from most of the sample area even above $T_\mathrm{c-bulk}$ 
though it contains a large fraction of the normal-state Sr$_2$RuO$_4$ regions. 
These results suggest that the anisotropy associated with the RuO$_2$ plane plays an essential role in the spatial development of the 3-K phase superconductivity.

We should here remind the fact that 
the in-plane \cite{Kambara2008PRL} as well as the out-of-plane \cite{Ando1999JPSJ} resistivity decrease by nearly 80\% even at 2.5~K from the normal-state value.
The observed shielding fraction, at most 1.2\% at 1.8~K for any field direction, 
seems to be apparently inconsistent with the decrease of the resistivity.
We will discuss the origin of this apparent inconsistency in Sec.~4.

\subsection{Field-magnitude dependence of $\chi_\mathrm{ac}$}

\begin{figure}
\begin{center}
\includegraphics[width=3in]{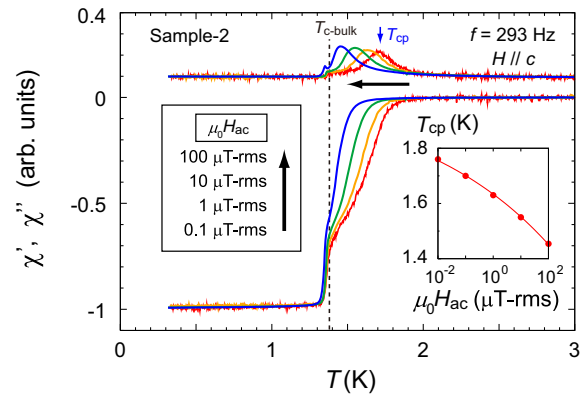}
\end{center}
\caption{(Color online) Ac magnetic field dependence of the real and imaginary parts of $\chi_\mathrm{ac}(T)$ of Sample-2 for $H_\mathrm{ac} \parallel c$ in zero dc field.
The inset represents the dependence of the peak temperature $T_\mathrm{cp}^\mathrm{ac}$ on the ac magnetic field amplitude.}
\label{acac}
\end{figure}

\begin{figure}
\begin{center}
\includegraphics[width=3in]{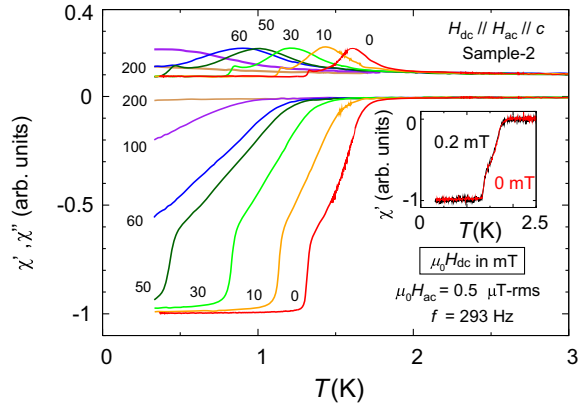}
\end{center}
\caption{(Color online) Dc magnetic field dependence of the real and imaginary parts of $\chi_\mathrm{ac}(T)$ of Sample-2 for $H_\mathrm{dc} \parallel H_\mathrm{ac} \parallel c$ ($\mu_0H_\mathrm{ac}=0.5$~$\muup$T-rms).
The numbers labeling the curves indicate the strength of the applied dc magnetic field $\mu_0H_\mathrm{dc}$ in mT.
The inset compares $\chi_\mathrm{ac}(T)$ in 0.2~mT with that in zero dc field.}
\label{acdc}
\end{figure}

In order to further investigate the development of the 3-K phase superconductivity,
we compare the $H_\mathrm{ac}$ and $H_\mathrm{dc}$ dependence of the $\chi_\mathrm{ac}(T)$ curve for $H_\mathrm{ac} \parallel H_\mathrm{dc} \parallel c$ using Sample-2.
We choose the values of $C_1$ and $C_2$ so that $\chi^\prime$ at 3~K becomes 0 and $\chi^\prime$ at 0.3~K becomes $-1$ for the 0.1~$\muup$T-rms curve;
we use the same $C_1$ and $C_2$ throughout the curves in Figs.~\ref{acac} and \ref{acdc}.
We first focus on the result for $\mu_0H_\mathrm{ac}=0.1$~$\muup$T-rms and $\mu_0H_\mathrm{dc}=0$~T (the red curve in Fig.~\ref{acac}).
Two anomalies were observed:
the sharp drop at 1.4~K (= $T_\mathrm{c-bulk}$) due to the bulk superconducting transition of Sr$_2$RuO$_4$ and 
the broad decrease between 1.4~K and 2~K corresponding to the 3-K phase superconductivity.
Just above $T_\mathrm{c-bulk}$, $\Delta \chi^\prime$($=\chi^\prime(\mathrm{3~K})-\chi^\prime(T)$) reaches 70\% of $\Delta \chi^\prime$ in the full Meissner state below 1.3~K.
Probably, $\Delta\chi^\prime$ just above $T_\mathrm{c-bulk}$ corresponds to the volume fraction of the 3-K phase in the samples.

We next give attention to the $H_\mathrm{ac}$ dependence of $\chi_\mathrm{ac}$ of the 3-K phase.
Figure~\ref{acac} represents the dependence on $H_\mathrm{ac}$ along the $c$ axis in zero dc field.
Interestingly, there is a strong $H_\mathrm{ac}$ dependence in $\chi_\mathrm{ac}$ except for the immediate vicinity of $T_\mathrm{c-bulk}$ as well as below $T_\mathrm{c-bulk}$.
In order to characterize the $H_\mathrm{ac}$ dependence,
we plot the peak temperature $T_\mathrm{cp}$ in $\chi^{\prime \prime}$ in the inset of the Fig.~\ref{acac}.
With increasing the amplitude of $H_\mathrm{ac}$, $T_\mathrm{cp}$ shifts toward lower temperatures.
As temperature approaches down to $T_\mathrm{c-bulk}$, 
$\chi^\prime$ for different $H_\mathrm{ac}$ apparently converge to the volume fraction of the 3-K phase.
Below $T_\mathrm{c-bulk}$, there is little $H_\mathrm{ac}$ dependence, as is expected for a bulk superconductivity.
We confirmed these features for $H \parallel ab$, too.

It is known that ``weak'' superconducting regions such as granular superconductors \cite{Yang1994PRB} respond sensitively to the ac field amplitude in measurements of $\chi_\mathrm{ac}$. 
Our observations of $H_\mathrm{ac}$ dependence in $\chi_\mathrm{ac}$ suggest that the 3-K phase superconductivity is a kind of ``weak'' superconductivity. 
We discuss details of this superconductivity in the next section.

In contrast with the strong $H_\mathrm{ac}$ dependence, the broad anomaly is relatively stable against $H_\mathrm{dc}$.
Figure~\ref{acdc} presents the temperature dependence of $\chi_\mathrm{ac}$ of Sample-2 in various $H_\mathrm{dc}$ along the $c$ axis.
In these measurements, we chose $\mu_0H_\mathrm{ac}$ of 0.5 $\muup$T-rms so that $H_\mathrm{ac}$ does not strongly suppress the broad anomaly (see Fig.~\ref{acac}). 
We found no difference between the $\chi_\mathrm{ac}$ curves for $H_\mathrm{dc}=0$ and 0.2~mT, as shown in the inset of Fig.~\ref{acdc},
though the broad anomaly in $\chi_\mathrm{ac}(T)$ is strongly suppressed by $\mu_0H_\mathrm{ac}$ of 0.1~mT-rms.
Remarkably, the broad anomaly is even more stable than the bulk transition in dc fields: 
Although the bulk transition is suppressed below 0.3~K in $\mu_0H_\mathrm{dc}$ larger than 60~mT,
the broad anomaly still persists above 1~K even in $\mu_0H_\mathrm{dc}$=100~mT. 
This result reflects the higher $H_\mathrm{c2}$ for the 3-K phase superconductivity than that for the bulk Sr$_2$RuO$_4$.\cite{Ando1999JPSJ}

\section{Discussion}

From our measurements of $\chi_\mathrm{ac}$ and $M$, we have revealed various features of the 3-K phase superconductivity:
(i) A large ac shielding fraction is observed just above $T_\mathrm{c-bulk}$.
(ii) The shielding fraction for any field direction is at most 1.2\% at 1.8~K, seemingly inconsistent with the decrease of resistivity.
(iii) The 3-K phase superconductivity above $T_\mathrm{c-bulk}$ has a strong anisotropy against the field direction
with respect to the Sr$_2$RuO$_4$ crystal axes rather than to the orientation of Ru lamellae:
The shielding fraction as well as the onset temperature of the observable diamagnetic signal is enhanced for $H \parallel c$ than for $H \parallel ab$.
(iv) Above $T_\mathrm{c-bulk}$, $\Delta\chi^\prime$ is more easily suppressed by $H_\mathrm{ac}$ than by $H_\mathrm{dc}$,
suggesting a formation of a ``weak'' superconductivity.
This feature is observed for both $H_\mathrm{ac} \parallel c$ and $H_\mathrm{ac} \parallel ab$.

In the following, we discuss the origin of these interesting features observed in the 3-K phase.

The first possible scenario is based on a model in which
there are many Ru lamellae enough to cover the whole sample 
if we were to see through the sample from the applied-field direction.
In other words, this scenario assumes that the large shielding fraction originates only from individual interface between Sr$_2$RuO$_4$ and Ru.
This scenario is similar to the scenario proposed for the origin of the superconductivity observed in the Sr$_3$Ru$_2$O$_7$ region cut out from the Sr$_3$Ru$_2$O$_7$-Sr$_2$RuO$_4$ eutectic system. \cite{Kittaka2008PRB}
In the Sr$_3$Ru$_2$O$_7$ region, Sr$_2$RuO$_4$ thin slabs embedded as stacking faults were observed with a transmission electron microscope \cite{Fittipaldi2008EPL} and 
it is suggested that these thin slabs become superconducting and contribute to the large shielding fraction.\cite{Kittaka2008PRB,Kittaka2009JPCS-1}
Although this scenario can explain the features (i) and (iv),
it is difficult to explain the feature (iii):
If this scenario were the case, the anisotropy of the shielding fraction would be governed by the geometry of the Ru inclusions rather than the crystal structure of Sr$_2$RuO$_4$.
Therefore, this scenario is not suitable for the 3-K phase.

The second possible scenario is that a Josephson network induced by the proximity effect of the interfacial superconductivity is formed and 
that the Josephson screening current connecting the lamellae produces the large shielding fraction.
In this scenario, we assume that the 3-K phase (probably $p$-wave)\cite{Mao2001PRL} superconductivity occurring at the interfaces penetrates into the normal-state Sr$_2$RuO$_4$ due to the proximity effect 
and forms Josephson-type weak links among different interfaces.
Indeed, a strong $H_\mathrm{ac}$ dependence of the $\chi_\mathrm{ac}(T)$ curve being similar to the feature (iv) 
has been observed in materials with such a Josephson network (e.g. Ref.~19). 
This scenario also seems to be consistent with the other present observations as discussed below.

\begin{figure}
\begin{center}
\includegraphics[width=3.4in]{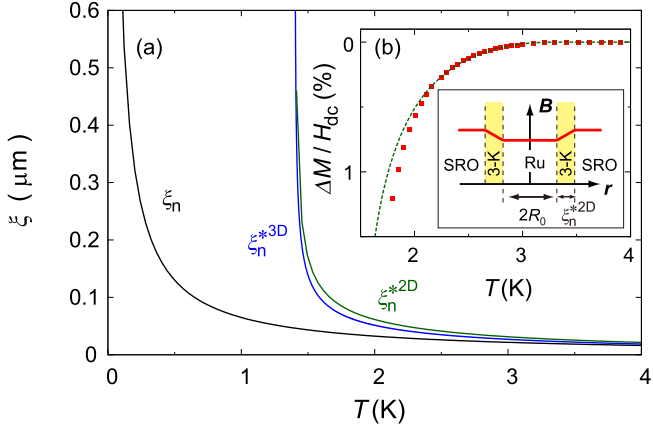}
\end{center}
\caption{(Color online) (a)Temperature dependence of the proximity length $\xi_\mathrm{n}$ and $\xi_\mathrm{n}^\ast$ calculated using eqs. \eqref{xi1}-\eqref{xi3} with $T_\mathrm{cn}$=1.4~K.
(b)$\Delta M/H_\mathrm{dc}$ in 2~mT for $H_\mathrm{dc} \parallel c$ (square symbol) and the result of the fitting with eq.~\eqref{M4} (dotted line).
The inset illustrates a model for our calculation.}
\label{xi}
\end{figure}

In the proximity-induced Josephson network scenario,
the observed large shielding fraction above $T_\mathrm{c-bulk}$ (feature (i)) indicates that
the proximity length should be comparable to or even larger than the inter-Ru distances (typically 10 $\muup$m).
The observation of the zero-voltage current above $T_\mathrm{c-bulk}$ \cite{Hooper2004PRB} also suggests a long proximity length. 
It is known that the proximity length $\xi_\mathrm{n}$ in a clean metal connected to a superconductor with its transition temperature $T_\mathrm{cs}$ is given by 
\begin{align}
\xi_\mathrm{n}(T)=\frac{\hbar v_\mathrm{F}}{2\pi k_\mathrm{B}T}\ \ \ , \label{xi1}
\end{align}
where $\hbar$ is the Dirac constant, $v_\mathrm{F}$ is the Fermi velocity of the normal metal, and $k_\mathrm{B}$ is the Boltzmann constant. 
If the normal metal is also a superconductor with a transition temperature $T_\mathrm{cn}$ lower than $T_\mathrm{cs}$, 
eq.~\eqref{xi1} is modified using the Eilenberger formalism:\cite{Delin1996SST} in the clean limit and 
in the temperature range $T_\mathrm{cn}<T<T_\mathrm{cs}$,
\begin{align}
\xi_\mathrm{n}^{\ast \mathrm{2D}}(T) = \xi_\mathrm{n}(T) \frac{1+2/\ln(T/T_\mathrm{cn})}{[1+4/\ln(T/T_\mathrm{cn})]^{1/2}} \label{xi2}
\end{align}
for a two-dimensional system, and 
\begin{align}
\frac{1}{2} \ln \frac{T}{T_\mathrm{cn}} =\frac{\xi_\mathrm{n}^{\ast \mathrm{3D}}(T)}{\xi_\mathrm{n}(T)}\mathrm{arctanh}\Biggl(\frac{\xi_\mathrm{n}(T)}{\xi_\mathrm{n}^{\ast \mathrm{3D}}(T)}\Biggl)-1 \label{xi3}
\end{align}
for a three-dimensional system.
In these equations, $\xi_\mathrm{n}(T)$ is given by eq.~\eqref{xi1}.
We plot in Fig.~\ref{xi} (a) the temperature dependence of $\xi_\mathrm{n}$ (diverging at 0~K) and $\xi_\mathrm{n}^\ast$ (diverging at $T_\mathrm{cn}$) calculated 
using $v_{\mathrm{F} \parallel ab}=5.5 \times 10^4$~m/s for the $\gamma$ band of Sr$_2$RuO$_4$ \cite{Bergemann2003AP} and $T_\mathrm{cn}=T_\mathrm{c-bulk}=1.4$~K.
Equations~\eqref{xi2} and \eqref{xi3} lead to the divergent behavior $\xi_\mathrm{n}^\ast(T) \propto v_\mathrm{F}/(T-T_\mathrm{cn})^{1/2}$ near $T_\mathrm{cn}$.
We note that $\xi_\mathrm{n}^\ast(T)$ in the dirty limit also exhibits $v_\mathrm{F}/(T-T_\mathrm{cn})^{1/2}$ behavior near $T_\mathrm{cn}$.\cite{Delin1996SST}
It is clear in Fig.~\ref{xi} (a) that the long-range proximity length discussed above for the normal-state Sr$_2$RuO$_4$ in the 3-K phase is not consistent with eq.~\eqref{xi1}, 
but can be explained by eq.~\eqref{xi2} or \eqref{xi3}.

Within this Josephson network scenario, the origin of the feature (iii), 
the anisotropy of the shielding fraction reflecting the layered crystal structure of Sr$_2$RuO$_4$,
is ascribable to the anisotropy of the proximity length $\xi_{\mathrm{n} \parallel ab}^\ast \gg \xi_{\mathrm{n} \parallel c}^\ast$
resulting from the fact $v_{\mathrm{F} \parallel ab} \gg v_{\mathrm{F} \parallel c}$.\cite{Mackenzie2003RMP}
Because of this anisotropy, 
the 3-K phase superconductivity penetrates more deeply into the normal-state Sr$_2$RuO$_4$ region along the $ab$ plane than along the $c$ axis.

Here, we discuss the origin of the feature (ii), the small shielding fraction above 2~K.
One possible scenario is that 
most of the interfaces become superconducting, leading to the large reduction in the resistivity, 
while these interfacial regions cannot prevent magnetic fields from penetrating into the Ru metal inside.
Let us calculate the temperature dependence of $M$ near the onset temperature with a model that 
Ru metal is cylindrical along the $c$ axis with the radius $R_0=1$ $\muup$m ($R_0 \gg \xi_\mathrm{n}^{\ast \mathrm{2D}}$) and 
quantized vortices are pinned only in the region $R_0<r<R_0+\xi_\mathrm{n}^{\ast \mathrm{2D}}(T)$.
(For simplicity, we represent by $\xi_\mathrm{n}^{\ast \mathrm{2D}}$ the region of 
both the interfacial superconductivity and the proximity effect.)
To this interfacial region, we apply the critical state model \cite{Thinkham2nd5-6}  
represented by the homogeneous critical current $J_\mathrm{c}(r,T)=J_\mathrm{c}(T)=-\mathrm{d}H(r)/\mathrm{d}r$,
as illustrated in the inset of Fig~\ref{xi} (b).
Then, the magnetic flux density $B(r,T)$ in the Ru metal is given by 
\begin{equation}
B(r,T)|_{r \le R_0}=\mu_0(H_\mathrm{dc}-J_\mathrm{c}(T)\xi_\mathrm{n}^{\ast \mathrm{2D}}(T)). \label{M1}
\end{equation}
We assume that $J_\mathrm{c}(T)$ behaves as $J_\mathrm{c}(0)(1-T/T_\mathrm{cs})^2$, 
adopting the same temperature dependence in the weak superconducting region in a granular superconductor.\cite{Muller1989PhysicaC} 
The diamagnetic magnetization per unit volume is expressed as
\begin{align}
\Delta M(T) = \frac{\langle B(T) \rangle}{\mu_0} -H_\mathrm{dc}. \label{M2}
\end{align}
Here, $\langle B(T) \rangle$ is the spatial average of the magnetic flux density in the bulk of the sample.
Because of the condition $R_0 \gg \xi_\mathrm{n}^{\ast \mathrm{2D}}$, $\langle B(T) \rangle$ is approximated as
\begin{align}
\langle B(T) \rangle \simeq f_\mathrm{Ru}B(R_0,T)+ (1-f_\mathrm{Ru})\mu_0H_\mathrm{dc}, \label{M3}
\end{align}
where $f_\mathrm{Ru}$ is the volume fraction of Ru inclusions in the sample. 
From eqs.~\eqref{M1}--\eqref{M3}, we obtain
\begin{equation}
\Delta M(T) \simeq f_\mathrm{Ru}J_\mathrm{c}(0)\biggl(1-\frac{T}{T_\mathrm{cs}}\biggl)^2\xi_\mathrm{n}^{\ast \mathrm{2D}}(T). \label{M4}
\end{equation}
Equation ~\eqref{M4} gives a good fitting to the data of $M$ for $H_{\mathrm{dc} \parallel c}=2$~mT above 2.1~K,
as plotted in Fig.~\ref{xi} (b).
The resultant fitting parameters are $f_\mathrm{Ru}J_\mathrm{c}(0)=25$~A/cm$^2$ and $T_\mathrm{cs}=3.3$~K.
If we adopt the ratio of total Ru area to the area of the $ab$ surface, approximately 0.05, as $f_\mathrm{Ru}$,
$J_\mathrm{c}(0)$ becomes 500 A/cm$^2$.
This value of $J_\mathrm{c}(0)$ seems reasonable
because it is comparable to the experimental value ($J_\mathrm{c}(0) \sim 800$ A/cm$^2$).\cite{Hooper2004PRB}
Below 2.1~K, the fitting is less satisfactory: the experimental shielding fraction is more pronounced than our calculation.
This deviation is attributable to the additional shielding by the inter-Ru supercurrent forming the Josephson network, 
which is not included in our calculation.

Summarizing the discussion above, we elucidate the spatial development of the 3-K phase superconductivity.
As presented in Fig.~\ref{FD} (b), we revealed that the onset temperature in $M$ for $H_\mathrm{dc} \parallel c$ is as high as 3.5~K.
It should be emphasized that  
$M$ for $H_\mathrm{dc} \perp c$ as well as the out-of-plane resistance \cite{Ando1999JPSJ} 
exhibit clear superconducting signals only below 3~K.
These results suggest that, as temperature decreases, 
the 3-K phase superconductivity above 3~K is confined within each RuO$_2$ plane at the Ru interface and is highly two dimensional.
When temperature further decreases, the 3-K phase superconductivity extends also along the $c$ axis.
At around 2.5~K, most of the interfacial regions exhibits non-bulk superconductivity, leading a large drop in resistivity.
This non-bulk superconductivity penetrates into the normal-state Sr$_2$RuO$_4$ region due to the proximity effect.
Below around 2~K, the proximity-induced Josephson network possibly develops connecting different interfaces and
extends more widely within the $ab$ plane than along the $c$ axis,
leading to the large and anisotropic shielding fraction.

According to the theoretical analysis by Sigrist and Monien,
the one-component $p$-wave order parameter with its node perpendicular to the interface 
would be stabilized at the interface near the 3-K onset temperature.
If weak links between different interfaces are formed with the one-component $p$-wave order parameter,
it is not possible that all links have zero phase shifts in the order parameter:
some links must have $\pi$ phase shifts.
This situation results in a frustration in the system because the links with the $\pi$ phase shift,
$\pi$-junctions, have slightly higher energy than the links with the zero phase shift.
In our ac-susceptibility study, however, we have not obtained any experimental hints indicating 
a formation of such a $\pi$-junction network, e.g.
paramagnetic signals associated with a spontaneous current in $\pi$-junction loops.
When the temperature is lowered, the other component is expected to emerge in the 3-K phase superconductivity 
leading to the chiral state $k_x+ik_y$.
Based on the observation of the zero-bias conductance peak, 
it has been suggested that the two-component state appears below about 2.3~K.\cite{Kawamura2005JPSJ}
Since our experimental results indicate that the Josephson network is formed mainly below 1.8~K,
it probably consists of the two-component $p$-wave order parameter in the chiral state $k_x+ik_y$. 
Whether or not the frustration occurs in the Josephson network 
needs to be further investigated both experimentally and theoretically.

\section{Conclusion}

We have clarified the process of spatial development of the 3-K phase superconductivity.
By precise magnetization measurements, we revealed that the onset temperature of the 3-K phase superconductivity is as high as 3.5~K. 
The observation of the diamagnetic shielding above 3~K only for $H_\mathrm{dc} \parallel c$ implies that
the shielding current originating from the 3-K phase superconductivity flows only within the RuO$_2$ planes.
Below 3~K, the diamagnetic signal associated with the shielding current along the $c$ axis also becomes observable. 
The small shielding fraction above around 2~K is explained by a scenario that 
the interfacial superconducting regions are too thin to exclude magnetic fields from Ru inclusions.
By ac susceptibility measurements, we revealed that the shielding fraction in the temperature range 
$T_\mathrm{c-bulk} < T < 2$~K is easily destroyed by $H_\mathrm{ac}$ and 
is anisotropic with respect to the crystal structure of Sr$_2$RuO$_4$.
In the vicinity of $T_\mathrm{c-bulk}$, the shielding fraction for all field directions reaches nearly 100\%.
Based on these experimental results, we suggest the formation of the Josephson network in the 3-K phase.
The proximity length diverging toward $T_\mathrm{c-bulk}$
enables the formation of weak links among well separated Ru metals.
Although the microscopic origin of the enhancement of $T_\mathrm{c}$ remains an open question,
the present results provide a basis for future theoretical and experimental studies. 

\acknowledgments
We thank Manfred Sigrist for valuable discussions. 
This work has been supported by the Grant-in-Aid for Global COE program ``The Next Generation of Physics, Spun from Universality and Emergence'' 
from the Ministry of Education, Culture, Sports, Science, and Technology (MEXT) of Japan.
It is also supported by Grants-in-Aid for Scientific Research from MEXT and from the Japan Society for the Promotion of Science (JSPS).
One of the authors (S. K.) is financially supported as a JSPS Research Fellow.


\end{document}